\documentclass[12pt,a4paper,final]{revtex4}

\usepackage{sidecap}

\usepackage{ulem}
\usepackage{epsfig}
\usepackage{amsmath,amssymb,amsthm,mathtools}
\usepackage{graphicx}
\usepackage{bm}
\usepackage{color,soul}

\DeclareMathAlphabet{\mathpzc}{OT1}{pzc}{m}{it}

\begin{document}

\renewcommand{\textfraction}{0.00}


\newcommand{\pT}{p_\perp}
\newcommand{\vAi}{{\cal A}_{i_1\cdots i_n}}
\newcommand{\vAim}{{\cal A}_{i_1\cdots i_{n-1}}}
\newcommand{\vAbi}{\bar{\cal A}^{i_1\cdots i_n}}
\newcommand{\vAbim}{\bar{\cal A}^{i_1\cdots i_{n-1}}}
\newcommand{\htS}{\hat{S}}
\newcommand{\htR}{\hat{R}}
\newcommand{\htB}{\hat{B}}
\newcommand{\htD}{\hat{D}}
\newcommand{\htV}{\hat{V}}
\newcommand{\cT}{{\cal T}}
\newcommand{\cM}{{\cal M}}
\newcommand{\cMs}{{\cal M}^*}
\newcommand{\vk}{\vec{\mathbf{k}}}
\newcommand{\bk}{\bm{k}}
\newcommand{\kt}{\bm{k}_\perp}
\newcommand{\kp}{k_\perp}
\newcommand{\km}{k_\mathrm{max}}
\newcommand{\vl}{\vec{\mathbf{l}}}
\newcommand{\bl}{\bm{l}}
\newcommand{\bK}{\bm{K}}
\newcommand{\bb}{\bm{b}}
\newcommand{\qm}{q_\mathrm{max}}
\newcommand{\vp}{\vec{\mathbf{p}}}
\newcommand{\bp}{\bm{p}}
\newcommand{\vq}{\vec{\mathbf{q}}}
\newcommand{\bq}{\bm{q}}
\newcommand{\qt}{\bm{q}_\perp}
\newcommand{\qp}{q_\perp}
\newcommand{\bQ}{\bm{Q}}
\newcommand{\vx}{\vec{\mathbf{x}}}
\newcommand{\bx}{\bm{x}}
\newcommand{\tr}{{{\rm Tr\,}}}
\newcommand{\sNN}{s_{\mathrm{NN}}}
\newcommand{\bc}{\textcolor{blue}}

\newcommand{\beq}{\begin{equation}}
\newcommand{\eeq}[1]{\label{#1} \end{equation}}
\newcommand{\ee}{\end{equation}}
\newcommand{\bea}{\begin{eqnarray}}
\newcommand{\eea}{\end{eqnarray}}
\newcommand{\beqar}{\begin{eqnarray}}
\newcommand{\eeqar}[1]{\label{#1}\end{eqnarray}}

\newcommand{\half}{{\textstyle\frac{1}{2}}}
\newcommand{\ben}{\begin{enumerate}}
\newcommand{\een}{\end{enumerate}}
\newcommand{\bit}{\begin{itemize}}
\newcommand{\eit}{\end{itemize}}
\newcommand{\ec}{\end{center}}
\newcommand{\bra}[1]{\langle {#1}|}
\newcommand{\ket}[1]{|{#1}\rangle}
\newcommand{\norm}[2]{\langle{#1}|{#2}\rangle}
\newcommand{\brac}[3]{\langle{#1}|{#2}|{#3}\rangle}
\newcommand{\hilb}{{\cal H}}
\newcommand{\pleft}{\stackrel{\leftarrow}{\partial}}
\newcommand{\pright}{\stackrel{\rightarrow}{\partial}}

\newcommand{\meqbox}[2]{\eqmakebox[#1]{$\displaystyle#2$}}

\newcommand{\Lcoll}{\frac{d E_{col}}{d \tau}}
\newcommand{\Ldndx}{\frac{d^2N_{\mathrm{rad}}}{dx d \tau}}
\newcommand{\Lnorm}{\frac{dN_{\mathrm{rad}}}{d \tau}}
\newcommand{\normtr}{\overline{N}_{tr}}
\newcommand{\dndxtr}{\overline{N}'_{tr}}
\newcommand{\dAtr}[1]{dA^{tr}_{#1}}
\newcommand{\Lrad}{R^{tr}_{\mathrm{rad}}}
\newcommand{\mucoll}{\overline{E}^{tr}_{\mathrm{col}}}
\newcommand{\collDist}{P^{tr}_{\mathrm{col}}}
\newcommand{\sigmacoll}{\sigma^{tr}_{\mathrm{col}}}
\newcommand{\dsdpti}{\frac{d^2 \sigma}{dp_{i}^2}}
\newcommand{\dsdptiRadL}{\frac{d^2 \sigma^{\mathrm{rad}}}{dp_{i}^2}}
\newcommand{\RAAtr}{R_{AA}^{tr}}


\title{DREENA-A framework as a QGP tomography tool}

\author{Dusan Zigic}
\affiliation{Institute of Physics Belgrade, University of Belgrade, Serbia}

\author{Igor Salom}
\affiliation{Institute of Physics Belgrade, University of Belgrade, Serbia}

\author{Jussi Auvinen}
\affiliation{Institute of Physics Belgrade, University of Belgrade, Serbia}

\author{Pasi Huovinen}
\affiliation{Institute of Physics Belgrade, University of Belgrade, Serbia}
\affiliation{Incubator of Scientific Excellence---Centre for Simulations of
             Superdense Fluids, University of Wroc\l{}aw, Poland}

\author{Magdalena Djordjevic\footnote{E-mail: magda@ipb.ac.rs}}
\affiliation{Institute of Physics Belgrade, University of Belgrade, Serbia}

\begin{abstract}
We present a fully optimised framework DREENA-A based on a
state-of-the-art energy loss model. The framework can include any, in
principle arbitrary, temperature profile within the dynamical energy
loss formalism. Thus, 'DREENA' stands for Dynamical Radiative and
Elastic ENergy loss Approach, while 'A' stands for Adaptive. DREENA-A
does not use fitting parameters within the energy loss model, allowing
it to fully exploit differences in temperature profiles which are the only
input in the framework. The framework applies to light and heavy
flavor observables, different collision energies, and large and smaller systems. This, together
with the ability to systematically compare data and predictions within
the same formalism and parameter set, makes DREENA-A a unique
multipurpose QGP tomography tool.
\end{abstract}

%
\maketitle

\section{Introduction}

QCD predicted that a new form of matter~\cite{Collins,Baym}---
consisting of quarks, antiquarks, and gluons that are no
longer confined---is created at extremely high energy
densities. According to the current cosmology, this new state of
matter, called Quark-Gluon Plasma (QGP)~\cite{QGP1,QGP2,QGP3,QGP4},
existed immediately after the Big Bang~\cite{Stock}. Today, QGP is
created in 'Little Bangs', when heavy ions collide at
ultra-relativistic energies~\cite{QGP2,QGP3}. Such collisions lead to
an expanding fireball of quarks and gluons, which thermalises to form
QGP; the QGP then cools down, and when the temperature reaches a
critical point, quarks and gluons hadronise.

Successful production of this exotic state of matter at the
Relativistic Heavy Ion Collider (RHIC) and the Large Hadron Collider
(LHC) allowed systematical testing of different models of QGP
evolution against experimental data. Up to now, it has been established that
QGP is formed at the LHC and RHIC experiments through two main
lines~\cite{QGP2,QGP3,Stachel} of evidence: {\it i)} by comparison of
low momentum ($\pT$) measurements with relativistic hydrodynamic
predictions, which implied that created QGP is consistent with the
description of a nearly perfect fluid~\cite{KolbHeinz,Romatschke,
  HeinzSnellings}, {\it ii)} by comparison of high-$\pT$
data~\cite{Adams,Adcox, Aad,Aamodt,Chatrchyan} with pQCD predictions,
which showed that high-$\pT$ partons (jets) significantly interact
with an opaque medium. Beyond this discovery phase, the current
challenge is to investigate the properties of this extreme form of
matter.

While high-$\pT$ physics had a decisive role in the QGP
discovery~\cite{QGP2}, it was rarely used for understanding the bulk
medium properties. On the other hand, low-$\pT$ observables do not
provide stringent constraints to all parameters of the models used to
describe the evolution of QGP, and thus leave some properties of QGP
badly constrained~\cite{Nagle,Niemi,Koop,Auvinen}. Thus, it is
desirable to explore QGP properties through independent theory
and data set. We argue that this is provided by jet energy loss and
high-$\pT$ data, complementing the low-$\pT$ constraints to QGP.

To use high-$\pT$ theory and data as a QGP tomography tool, it is
necessary to have a realistic high-$\pT$ parton energy loss model. We
previously showed~\cite{MD_PLB,DDB_PLB,MD_5TeV,MD_PRL} that the
dynamical energy loss formalism provides a reliable tool for such
tomography. This formalism has the following properties necessary for
inferring the bulk QGP medium parameters: {\it i)} It is based on
finite size, finite temperature field theory~\cite{Kapusta,Le_Bellac},
and takes into account that QGP constituents are dynamical (moving)
particles. Consequently, all divergences are naturally regulated in
the model. {\it ii)} Both collisional~\cite{MD_Coll} and
radiative~\cite{MD_PRC,DH_PRL} energy losses are calculated in the
same theoretical framework. {\it iii)} It is applicable to both light
and heavy flavors, so it can provide predictions for an extensive set
of probes. {\it iv)} Temperature is a natural variable in the
framework~\cite{SIDD}, so that the $T$ profiles resulting from bulk medium
simulations are a direct input in the model. {\it v)} The
non-perturbative effects related to screening of the chromo-magnetic
and chromo-electric fields are included~\cite{MD_MagnMass} so that the
model can also capture the non-perturbative medium-related
interactions. {\it vi)} No fitting parameters are used in comparing
the dynamical energy loss predictions with high-$\pT$
data~\cite{Stojku:2020wkh,SAHD}, i.e., all
the parameters have been fixed to the standard literature values. For the
bulk medium tomography, this allows concentrating only on bulk medium
simulation parameters. While other available energy loss models (see
e.g.~\cite{BDMPS,BZ, ASW,GLV,AMY}) have some of the above properties,
none have all (or even most of them), making the dynamical energy loss
a unique framework for QGP tomography.

Including full medium evolution in the dynamical energy loss is,
however, a highly non-trivial task, as all the model properties have
to be preserved~\cite{BD_JPG}, without additional simplifications in
the numerical procedure. Furthermore, to be effectively used as a
precision QGP tomography tool, the framework needs to efficiently
(timewise) generate a comprehensive set of light and heavy flavor
suppression predictions through the same numerical framework and the
same parameter set. Such predictions can then be compared with the
available experimental data, sometimes even repeatedly (i.e., iteratively) -- for different combinations of QGP medium parameters -- to extract medium properties that are consistent with both low and high-$\pT$ data.

To introduce the medium evolution in the dynamical energy loss, we
took a step-by-step approach, allowing us to check the
consistency of each consecutive step by comparing its results with the previous (simpler) framework versions. Consequently, we first developed the DREENA-C
framework~\cite{DREENA-C} ('C' stands for constant temperature),
continuing to DREENA-B~\cite{DREENA-B} ('B' stands for Bjorken
expansion). In this manuscript, we present a fully optimised
DREENA-A framework, where 'A' stands for 'adaptive' (i.e., arbitrary)
temperature evolution. The convergence speed of the developed
numerical procedure is analysed, as well as consistency with other
(earlier) versions of the framework, as necessary for the reliable and
efficient QGP tomography tool. Finally, as a utility check of the
DREENA-A framework, the sensitivity of high-$\pT$ observables to
different temperature profiles is presented.

\section{Theoretical outline}

We use the generic pQCD convolution to calculate the final quenched and unquenched spectra
of hadrons:
\begin{eqnarray}
\frac{E_f d^3\sigma_q(H_Q)}{dp_f^3} = \frac{E_i d^3\sigma(Q)}{dp^3_i}
 \otimes
{P(E_i \rightarrow E_f )}
\otimes D(Q \to H_Q)\, ,
\label{schem} \end{eqnarray}
\begin{eqnarray}
\frac{E_f d^3\sigma_u(H_Q)}{dp_f^3} = \frac{E_i d^3\sigma(Q)}{dp^3_i}
\otimes D(Q \to H_Q)\, .
\label{schem2} \end{eqnarray}
$\frac{E_f d^3\sigma_q(H_Q)}{dp_f^3}$ is the final hadron spectrum in the presence of QGP, while $\frac{E_f d^3\sigma_u(H_Q)}{dp_f^3}$ is the spectrum in the absence of QGP. '$i$' and '$f$' correspond to 'initial' and 'final', respectively. $Q$ denotes quarks and gluons, while $H_Q$ denotes hadrons. Initial parton spectrum
is denoted by $E_i d^3\sigma(Q)/dp_i^3$, and computed at
next to leading order~\cite{Vitev0912,Cacciari} for light and heavy partons.
$P(E_i \rightarrow E_f )$ is the probability for energy transfer, which
includes medium induced radiative~\cite{MD_PRC,DH_PRL} and
collisional~\cite{MD_Coll} contributions in a finite size dynamical QCD
medium with running coupling~\cite{MD_PLB}. Both contributions include multi-gluon fluctuations, introduced according to Refs.~\cite{GLV_suppress,MD_PLB} for
radiative and~\cite{Moore:2004tg,WHDG} for collisional
energy loss (for more details, see below). $Q$ to hadron $H_Q$ fragmentation is denoted by $D(Q \to
H_Q)$. For charged hadrons we use DSS~\cite{DSS}, for D mesons BCFY~\cite{BCFY} and for B mesons KLP~\cite{KLP} fragmentation functions, respectively.

In DREENA-A, the medium temperature needed to calculate $P(E_i \rightarrow E_f )$
depends on the position of the parton according to a temperature profile
given as an input. Therefore, the temperature that the parton experiences
along its path, becomes a function of the coordinates of its origin ($\mathrm{x}_0, \mathrm{y}_0$),
the angle of its trajectory $\phi$, and the
proper time $\tau$:
\begin{equation}
  T(\mathrm{x}_0, \mathrm{x}_0, \phi, \tau)
    = T_{profile}(\mathrm{x}_0 + \tau \cos \phi, \mathrm{y}_0 + \tau \sin \phi, \tau),
  \label{TemperatureAlongPath}
\end{equation}
where $T_{profile}$ is, in principle, arbitrary. This temperature then appears in the expressions below.

The collisional energy loss is given by the following analytical
expression~\cite{DREENA-B}:
\beqar
\frac{d E_{col}}{d \tau} &=& \frac{2 C_R}{\pi \, v^2} \alpha_S (E \, T) \, \alpha_S (\mu^2_E(T))\times \nonumber\\
&& \hspace*{-1.5cm}
\int_0^\infty n_{eq}(|\vec{\mathbf{k}}|, T) d |\vec{\mathbf{k}}| \;
\left( \int_0^{|\vec{\mathbf{k}}|/(1+v)} d |\vec{\mathbf{q}}|
\int_{-v |\vec{\mathbf{q}}|}^{v |\vec{\mathbf{q}}|}\; \omega d \omega \;+
\int_{|\vec{\mathbf{k}}|/(1+v)}^{|\vec{\mathbf{q}}|_{max}} d |\vec{\mathbf{q}}|
\int_{|\vec{\mathbf{q}}|-2|\vec{\mathbf{k}}| }^{v |\vec{\mathbf{q}}|}\;
\omega d \omega \; \right)\times \nonumber \\
&& \hspace*{-1.5cm}\left( |\Delta_L(q,T)|^2 \frac{(2 |\vec{\mathbf{k}}|+\omega)^2
- |\vec{\mathbf{q}}|^2}{2}  +
|\Delta_T(q,T)|^2 \frac{(|\vec{\mathbf{q}}|^2-\omega^2)
((2 |\vec{\mathbf{k}}|+\omega)^2+ |\vec{\mathbf{q}}|^2)}
{4 |\vec{\mathbf{q}}|^4} (v^2 |\vec{\mathbf{q}}|^2-\omega^2) \right).
\eeqar{Eel_infinite}
Here we used the following notation: $k$ is the 4-momentum of the
incoming medium parton; $T$ is the current temperature along the path,
given by Eq.\ (\ref{TemperatureAlongPath});
$n_{eq}(|\vec{\mathbf{k}}|,T)
 =\frac{N}{e^{|\vec{\mathbf{k}}|/T}-1}+\frac{N_f}{e^{|\vec{\mathbf{k}}|/T}+1}$
is the equilibrium momentum distribution~\cite{BT} at temperature $T$
including quarks and gluons ($N$ and $N_f$ represent, respectively,
the number of colors and flavors); $v$ denotes velocity of the
incoming jet; $q=(\omega, \vec{\mathbf{q}})$ is the 4-momentum of the
exchanged gluon; $E=p^2+M^2$ denotes the initial jet energy, $p$ is the jet momentum,  while $M$ is the mass of the quark or gluon jet;
$C_R=\frac{4}{3}$ for quark jet and $3$ for gluon jet;
$\Delta_L (T)$ and $\Delta_T (T)$ are effective longitudinal and
transverse gluon propagators~\cite{Gyulassy_Selikhov}, while the
electric screening (the Debye mass) $\mu_E(T)$ is obtained by
self-consistently solving the expression from~\cite{Peshier} ($\Lambda_{QCD}$ is perturbative QCD scale):
\beqar
\frac{\mu_E(T)^2}{\Lambda_{QCD}^2} \ln \left(\frac{\mu_E(T)^2}{\Lambda_{QCD}^2}\right)=\frac{1+N_f/6}{11-2/3 \, N_f} \left(\frac{4 \pi T}{\Lambda_{QCD}} \right)^2.
\eeqar{mu}
Running coupling $\alpha_S (Q^2)$ is defined as~\cite{Field}
\beqar
\alpha_S (Q^2)=\frac{4 \pi}{(11-2/3 N_f) \ln (Q^2/\Lambda_{QCD}^2)},
\eeqar{alpha}
where, in the collisional  energy loss case, the coupling appears through the term
$\alpha^2_S$~\cite{MD_Coll}, which can be factorised to $\alpha_S (\mu_E^2) \, \alpha_S (E \, T)$~\cite{Peigne2008} (see also~\cite{MD_PLB}).

The radiation spectrum~\cite{DREENA-B} is:
\beqar
\Ldndx &=&
 \int \frac{d^2k}{\pi} \,\frac{d^2q}{\pi} \, \frac{2\, C_R  C_2(G)\, T}{x} \,  \frac{\mu_E (T)^2 -\mu_M(T)^2 }{(\bq^2+\mu_M (T)^2) (\bq^2+\mu_E (T)^2)} \, \frac{\alpha_S (E \, T) \, \alpha_S (\frac{\bk^2+\chi (T)}{x}) }{\pi}\nonumber\\
    && \hspace*{-1cm} \times\frac{(\bk{+}\bq)}{(\bk{+}\bq)^2+\chi(T)} \left(1-\cos{\left(\frac{(\bk{+}\bq)^2+\chi(T)}{x E^+} \, \tau\right)}\right)
        \left(\frac{(\bk{+}\bq)}{(\bk{+}\bq)^2+\chi(T)}
    - \frac{\bk}{\bk^2+\chi(T)}
    \right).
\eeqar{DeltaNDynTau}

Here $C_2(G)=3$; $\chi (T) \equiv M^2 x^2 + m_g (T)^2$, where $x$ is
the longitudinal momentum fraction of the jet carried away by the
emitted gluon, and
$m_g (T)= \mu_E(T)/\sqrt 2$ is the effective gluon mass in finite
temperature QCD medium~\cite{DG_TM}; $M=1.2$~GeV for charm, 4.75~GeV for bottom and $\mu_E(T)/\sqrt 6$ for light quarks; $\mu_M (T)$ is magnetic
screening~\cite{Maezawa,Nakamura}; $\bq$ and $\bk$ are transverse
momenta of exchanged (virtual) and radiated gluon, respectively.
$Q_k^2=\frac{\bk^2 +\chi (T)}{x}$ in $\alpha_S (\frac{\bk^2+\chi (T)}{x})$ corresponds to the off-shellness of the jet prior to the gluon radiation~\cite{MD_PRC}. Note that, all $\alpha_S$ terms in Eqs.~(\ref{Eel_infinite}) and (\ref{DeltaNDynTau}) are infrared safe (and moreover of a moderate value)~\cite{MD_PLB}. Thus, contrary to majority of other approaches, we do not need to introduce a cut-off in $\alpha_S (Q^2)$.

We further assume that radiative and collisional energy losses are small, i.e., much smaller than initial jet energy, so that these contributions can be separately treated in $P(E_i \rightarrow E_f )$, i.e., jet quenching is performed via two independent branching processes~\cite{MD_PLB,WHDG}.

To obtain the radiative energy loss contribution to the suppression~\cite{GLV_suppress}, we start with Eq.~(\ref{DeltaNDynTau}) and, for a given trajectory, we first compute the mean number of gluons emitted due to induced radiation  (further denoted as $\normtr(E)$), as well as the mean number of gluons emitted per fractional energy loss $x$ (i.e., $\frac{d\overline{N}_{tr}(E)}{dx}$, for compactness further denoted as $\dndxtr(E, x)$):
\beqar
\normtr(E)  =\int\displaylimits_{tr} \left( \int \Ldndx dx \right) d\tau, \qquad
\dndxtr(E, x) = \int\displaylimits_{tr} \Ldndx d\tau,
\eeqar{dndxtr}
where the subscript $tr$ indicates that the value depends on the trajectory. Radiative energy loss suppression takes multi-gluon fluctuations into account and, if we assume that the fluctuations of gluon number are uncorrelated, the radiative energy loss probability can be expressed via Poisson expansion~\cite{GLV_suppress,MD_PLB}:
\beqar
P^{tr}_{rad}(E_i \rightarrow E_f ) &=& \frac{\delta(E_i-E_f)}{e^{\normtr(E_i)}}+\frac{\dndxtr(E_i,1-\frac{E_f}{E_i})}{e^{\normtr(E_i)}}+ \nonumber \\
&&\hspace*{-2.5cm}+\sum_{n=2}^{\infty}\frac{e^{-\normtr(E_i)}}{n!}\!\int \!dx_1 \!\cdots \!dx_n \dndxtr(E_i,x_1) \!\cdots\! \!\dndxtr(E_i,x_{n-1})\dndxtr(E_i,1\!-\!\frac{E_f}{E_i}\!-\!x_1\!-\!\cdots\!-\!x_{n-1}),
\eeqar{probability}
$E_i$ and $E_f$ are initial and final jet energy (before and after) radiative process.

To calculate the parton spectrum after radiative energy loss, we apply
\beqar
\frac{E_{f,R} d^3\sigma}{dp_{f,R}^3} = \frac{E_i d^3\sigma(Q)}{dp^3_i}
 \otimes
{P^{tr}_{rad}(E_i \rightarrow E_{f,R} )},
\eeqar{sigmaRad}
where the final spectra is obtained after integrating over $p_i>p_f$.

To find collisional energy loss contribution, Eq.~(\ref{Eel_infinite}) is first integrated over the given trajectory:
\beqar
\mucoll(E)  = \int\displaylimits_{tr} \frac{d E_{col}}{d \tau} d\tau.
\eeqar{colltr}
For collisional energy loss, the full fluctuation spectrum is approximated by a Gaussian centered at the average energy loss $\mucoll(E)$~\cite{Moore:2004tg,WHDG}
\beqar
\collDist(E_{i},E_{f})=\frac{1}{\sqrt{2\pi}\sigmacoll(E_i)} \exp\Big(-\frac{(E_i-E_f-\mucoll(E_i))^2}{2 \, \sigmacoll(E_i)^2}\Big),
\eeqar{collGauss}
with a variance
\beqar
\sigmacoll(E)=\sqrt{2\overline{T^{tr}}\,\mucoll(E)},
\eeqar{musigmaColl}
where $\overline{T^{tr}}$ is the average temperature along the trajectory, $E_i$ and $E_f$ are initial and final energy (before and after) collisional processes.

To calculate the quenched hadron spectrum after collisional energy loss, we apply
\beqar
\frac{E_f d^3\sigma_q (H_Q)}{dp_{f}^3} = \frac{E_{i,C} d^3\sigma(Q)}{dp^3_{i,C}}
 \otimes
{P^{tr}_{col}(E_{i,C} \rightarrow E_{f} )} \otimes D(Q \to H_Q)\, ,
\eeqar{sigmaRC}
where we assume $E_{i,C}=E_{f,R}$, i.e. the final jet energy after radiative quenching corresponds to the initial jet energy for collisional quenching. Since both collisional energy loss and gain contribute to the final spectra~\cite{MD_Coll, WHDG}, both $E_{i,C}>E_f$ and $E_{i,C}<E_f$ have to be taken into account in Eq.~(\ref{sigmaRC}). Finally, the hadron suppression $\RAAtr(p_f,H_Q)$ for the single trajectory, after radiative and collisional energy loss, is equal to the ratio of quenched and unquenched momentum spectra
\beqar
\RAAtr(p_f,H_Q) = \frac{E_f d^3\sigma_q(H_Q)}{dp_f^3}/\frac{E_f d^3\sigma_u(H_Q)}{dp_f^3},
\eeqar{RAAtr}
where $\frac{E_f d^3\sigma_u(H_Q)}{dp_f^3}$ is given by Eq.~(\ref{schem2}).
$\RAAtr(p_f,H_Q)$ then needs to be averaged over trajectories with the same direction angle $\phi$ to obtain the suppression as a function of angle, $R_{AA}(p_f, \phi,H_Q)$. This is an important intermediary step since, depending on the details of QGP temperature evolution and the spatial variations in the temperature profile, energy loss may significantly depend on the parton's direction of motion~\footnote{In earlier DREENA frameworks, this dependence was also present but was solely a consequence of the path-length distribution dependence on the angle.}. Once we have calculated $R_{AA}(p_f, \phi,H_Q)$, we can easily evaluate $R_{AA}$ and $v_2$ observables as~\cite{Luzum} (we here omit $H_Q$ in the expressions, and denote $p_f=p_\perp$):
\begin{equation}
 R_{AA}(p_\perp) = \frac{1}{2 \pi} \int_0^{2\pi} R_{AA}(p_\perp, \phi) d\phi, \label{RAAdefinition}
\end{equation}

\begin{equation}
 v_2(p_\perp) = \frac{\frac{1}{2 \pi} \int_0^{2\pi} \cos(2\phi) R_{AA}(p_\perp, \phi) d\phi }{R_{AA}(p_\perp)}.  \label{v2definition}
\end{equation}

While the general expressions of the dynamical energy loss formalism are the same as in the DREENA-B framework~\cite{DREENA-B}, the fact that, in DREENA-A, the temperature entering the Eqs.~(\ref{Eel_infinite}-\ref{DeltaNDynTau}) explicitly depends on the current parton position, notably complicates the implementation of these formulas, as we discuss in the following section.

\section{Framework development}

Our previous DREENA-C and DREENA-B frameworks were based on
computationally useful, but rough, approximations of the medium
evolution: while in DREENA-C, there was no evolution, and the temperature remained constant both in time and along spatial
dimensions, in DREENA-B, the medium was assumed to evolve according to
1D Bjorken approximation~\cite{BjorkenT}. Due to these approximations,
parton energy loss depended on its path length independently of its
direction or production point. This allowed to analytically integrate
energy-loss formulas to a significant extent, which notably reduced
the number of required numerical integrations. Furthermore $R_{AA}$
only needed to be averaged out over precalculated path-length
distributions. Thus, these approximations of the medium evolution
straightforwardly led to efficient computational algorithms for
DREENA-C and DREENA-B.

DREENA-A framework, on the other hand, addresses fully general medium
dynamics, with arbitrary spatio-temporal temperature
distribution. The main input to the algorithm is the temperature profile $T_{profile}$ given as a three-dimensional matrix of temperature values at points with coordinates $(\mathrm{x},\mathrm{y}, \tau)$ (in the input file, the values should be arranged in an array of quartets of the form $(\tau, \mathrm{x},\mathrm{y}, T_{profile})$, and the lowest value of $\tau$ appearing in the data is taken to be $\tau_0$). In addition to the temperature profile, the DREENA-A algorithm also takes, as inputs, the initial parton $p_\perp$ distributions $\frac{d^2 \sigma}{dp^2_\perp}$ (each as an array of $(p_\perp, \frac{d^2 \sigma}{dp^2_\perp})$ pairs) and the jet production probability distribution (as a matrix of probability density values in the transversal plane, formatted analogously as the profile temperature values). This level of generality requires a different approach
than in previous frameworks. Since the DREENA-A algorithm takes
arbitrary medium temperature evolution as the input, the energy loss has to be individually calculated for each parton trajectory.

This means that for each trajectory -- given by the coordinates $\mathrm{x}_0$ and $\mathrm{y}_0$ of the parton origin (in the transversal plane) and the direction angle $\phi$ -- we must first numerically evaluate integrals (\ref{dndxtr}) and (\ref{colltr}). Since the current parton position -- for a given trajectory -- becomes a function of the proper time $\tau$, integrands in (\ref{dndxtr}) and (\ref{colltr}) also become functions of $\tau$, either through an explicit dependence, or via position and time dependent medium temperature (\ref{TemperatureAlongPath}). We numerically integrate these functions along the trajectory (parametrized by $\tau$ as $\mathrm{x} = \mathrm{x}_0 + \tau \cos \phi$, $\mathrm{u} = \mathrm{y}_0 + \tau \sin \phi$), starting from the origin at $(\mathrm{x}_0, \mathrm{y}_0)$ and moving in small integration steps along the direction $\phi$ (in practice, $0.1$ fm step is sufficiently small for most of the profiles). The integration is terminated when the medium temperature at the current parton's position drops below $T_c = 155 MeV$~\cite{Tcritical}, i.e., when the parton leaves the QGP phase. Also, we approximate that there are no energy losses before the initial time $\tau_0$ (which is a parameter of the temperature profile evolution) and thus the first part of the trajectory, corresponding to $\tau < \tau_0$, is effectively skipped (i.e., $\tau_0$ is taken as the lower limit of integration in (\ref{dndxtr}) and (\ref{colltr})).

Once we, for a given trajectory, compute the integrals (\ref{dndxtr}) and (\ref{colltr}), we then perform the rest of procedure laid out by Eqs.\ (\ref{dndxtr}-\ref{RAAtr}). Most of the computation time is spent on numerical integrations, in particular for evaluating integrals in Eqs.~(\ref{probability},\ref{sigmaRad}). While, in principle, $n \rightarrow \infty$ in Eq.~(\ref{probability}), in practice we show that $n=5$ is sufficient for convergence in the case of quark jets, while for gluon jets $n=7$ is needed. In general, the Quasi-Monte Carlo integration method turned out to be the most efficient and is used for all these integrals (as quasirandom numbers, we use precalculated and stored Halton sequences). The result of the integration (\ref{RAAtr}) is the final hadron suppression $\RAAtr(p_\perp,H_Q)$ for the jet moving along the chosen trajectory, given as the function of its transversal momentum.

To obtain $R_{AA}(p_\perp, \phi,H_Q)$, we have to average this result over all production points (taking into account the provided jet production probability distribution) and repeat the procedure for many angles $\phi$. In practice, this means that we must evaluate energy loss along a very large number of trajectories.
This has significantly increased the computational complexity of the problem compared to DREENA-C and DREENA-B and required a number of optimisations.

\subsection{Numerical optimisation of DREENA-A}
\begin{figure*}
\centering
\epsfig{file=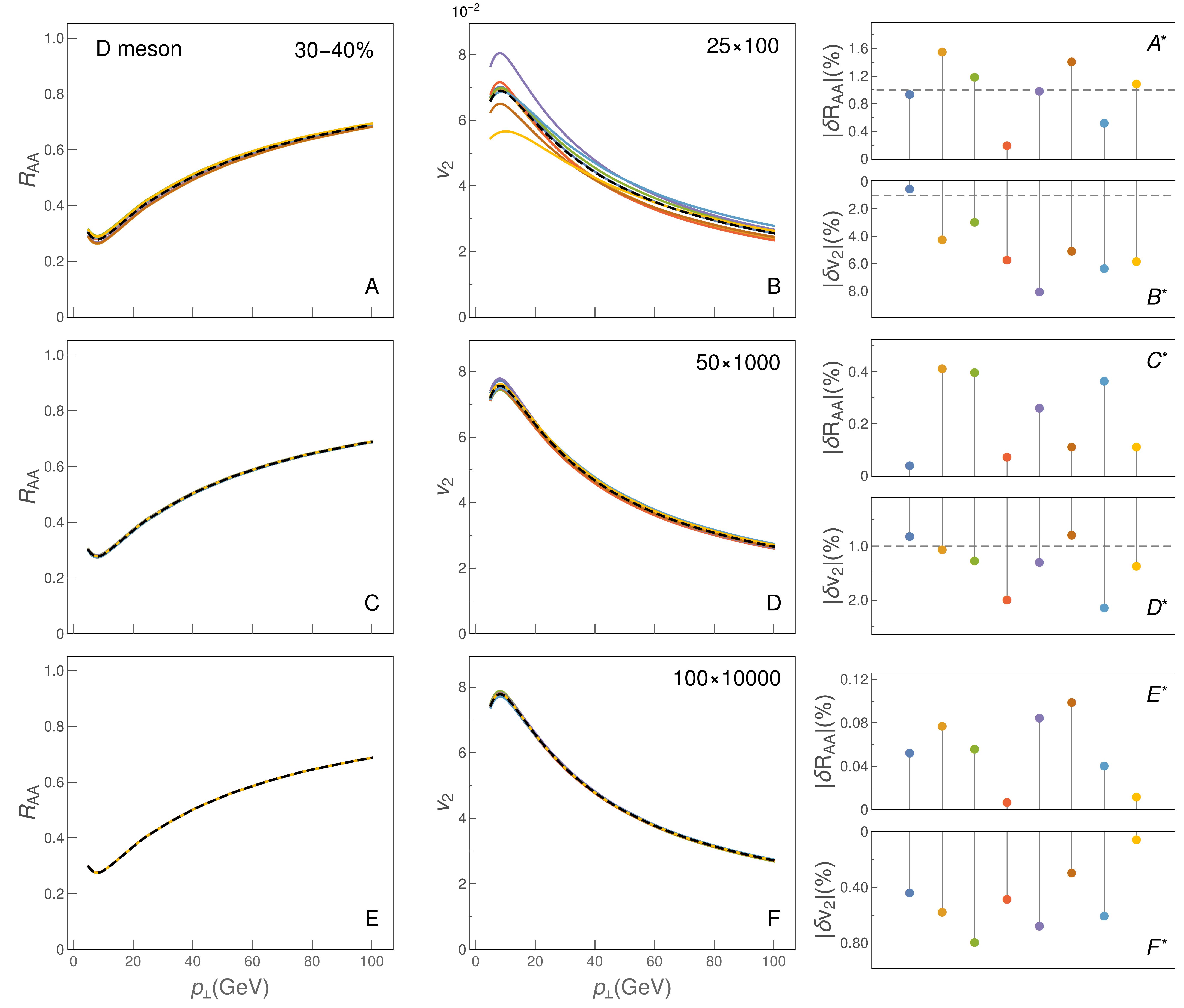,scale=0.375}
\vspace*{-0.2cm}
 \caption{ D meson $R_{AA}$ (left) and $v_2$ (middle) at 30-40$\%$
   centrality computed using different numbers of randomly generated
   trajectories (Monte Carlo approach), together with their deviations
   (right, scaled 1-norm was used as a metric) from the results averaged
   over the same ensemble of trajectories. The
   dashed horizontal line in rightmost panels indicates the threshold
   of 1$\%$ deviation. The top row depicts results obtained from sampling
   25 trajectories at different angles originating from each of 100
   randomly selected jet-production points; the middle row---50 angles
   from 1000 points; the bottom row---100 angles from 10000 points. Each
   panel shows the results of eight repeated computations (each with
   an independent ensemble of randomly generated trajectories), the dashed
   line representing the average. $M=1.2$~GeV,
   $\mu_M/\mu_E=0.5$~\cite{Maezawa,Nakamura}.}
\label{fig:MCConvergence}
\end{figure*}

We started by adapting optimisation methods that we successfully
implemented in earlier versions. One useful approach was a tabulation
and consequent interpolation of values for computationally expensive
functions. In particular, this is crucial for the complicated integrals (\ref{Eel_infinite}-\ref{DeltaNDynTau}): while a two dimensional array is sufficient to tabulate $\Lcoll$ (which is a function of $T$ and $p$), values of $\Ldndx$ (depending on $\tau, T, p$ and $x$) must be stored in a four-dimensional array. Tabulating such functions is done adaptively, with the density of evaluated points varying depending on the function behaviour (i.e., using a denser grid where the functions change rapidly and sparser where the behaviour is smooth). In the case of these two functions, not only that the consequent interpolation can significantly reduce the overall number of integral evaluations, but the corresponding tables (for each particle type) can be evaluated only once and then permanently stored and reused for all trajectories and even for different temperature profiles. To further optimise the algorithm, we also precalculate the integral $\int \Ldndx dx$ values and store a corresponding three-dimensional array (since it is a function of $\tau, T$, and $p$).


When using this table-interpolation method, it is often necessary to make a function transformation before tabulation: e.g., it is more efficient and accurate to sample and later interpolate logarithm of a rapidly (nearly or approximately-exponentially) increasing function than the function
itself (similarly, it is sometimes more optimal to tabulate ratio, or
a product of functions than each of the functions separately). For example, it is much more optimal to tabulate and consecutively interpolate $R_{AA}$s (and other similarly behaving expressions) than the corresponding momentum distributions. This methodology is now extensively applied throughout DREENA-A (from some intermediate-level energy loss results to evaluating multi-dimensional integrals in the calculation of radiated gluon rates). Given the size of some of these tables and that many interpolations are needed, we ensured that the table lookup and interpolation algorithm are efficient.

As we encounter multiple numerical integrations at different stages of the computation, modifying their order was another type of optimisation, where the natural order (from the theoretical viewpoint) is not necessarily followed but is instead adapted to the particular function behaviour. Specifically, it turned out that a different order of integration (for radiative contribution) is optimal for heavy flavor particles compared to gluons. I.e., while it is natural, from the physical perspective, to start with the initial momentum distributions of partons and integrate over the radiative energy loss (see Eqs.~(\ref{probability},\ref{sigmaRad})), it turned out that (for heavy flavors) the shape of the initial distributions necessitates a very high number of integration points to achieve the required computation precision. Reorganising the formulas and postponing the integration over initial distributions to the very end turned much more computationally optimal for heavy flavor. A similar procedure in the case of light quarks allowed much of the integration to be carried out jointly for all quarks, since their effective masses are the same, but initial $p_\perp$ distributions differ. Overall, this type of optimisation led to four-time faster execution times.

\begin{figure*}
\centering
\epsfig{file=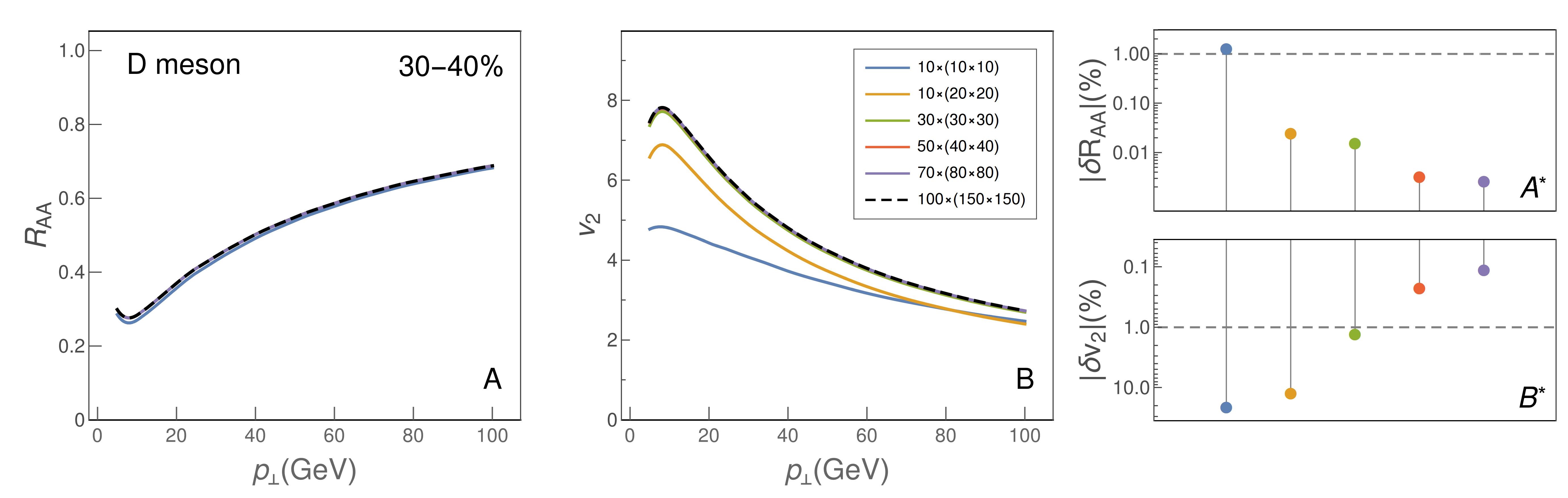,scale=0.375}
\vspace*{-0.2cm}
\caption{D meson $R_{AA}$ (left) and $v_2$ (middle) at 30-40$\%$
  centrality computed using different numbers of trajectories originating from equidistant points. Results are labeled by
  numbers $n_\phi \times (n_x \times n_y)$: jet directions are along
  $n_\phi$ uniformly distributed angles (from $0$ to $2 \pi$) originating
  from each point of the $n_x$-by-$n_y$ equidistant grid. Deviation of
  each line from the baseline result (chosen as the outcome for $100
  \times (150 \times 150)$ trajectories, dashed line) is shown in
  right panels. $M=1.2$~GeV, $\mu_M/\mu_E=0.5$.}
\label{fig:SCANConvergence}
\end{figure*}

The crucial optimisation in DREENA-A is the method used for averaging
over the particle trajectories. In suppression calculations,
it is common to carry out the averaging over production points
and directions by Monte Carlo (MC) sampling, but it turned out that the equidistant sampling of both jet production points and direction angles was here significantly more efficient.
We initially implemented the Monte Carlo
approach, randomly selecting both the origin coordinates and the
angles of particle trajectories. The binary collision density was used
as the probability density for coordinates of origins, while the
angles were generated from a uniform distribution. Convergence of the
results by using this method required a large number of sampled
trajectories, as illustrated in Fig.~\ref{fig:MCConvergence}. The
figure shows $R_{AA}$ and $v_2$ results obtained by the DREENA-A
algorithm for a different total number of trajectories (the
computation was done for D meson traversing the temperature evolution
generated using a Glauber initialised viscous hydrodynamic
code~\cite{Molnar:2014zha}, at 30-40\% centrality class). The plots in
the right column of Fig.~\ref{fig:MCConvergence} show the magnitude
of the deviation of the particular curve from the median curve,
where the latter is the arithmetic mean of all curves in the plot (as the measure of deviation of a function $f(p)$ from a reference function $\overline f(p)$ we use $|\delta f| = \frac{\int |f(p) - \overline f(p)| dp}{\int |\overline f(p)| dp}$). We see that
$R_{AA}$ convergence is easily achieved, where relative deviations
of the order of ~1\% are obtained by taking into account only 2500 trajectories (see
Fig.~\ref{fig:MCConvergence}-$A$ and Fig.~\ref{fig:MCConvergence}-$A^*$). Computing the $v_2$ value requires
much more trajectories, i.e., we see a substantial scattering of the
Monte Carlo results with 2500 trajectories, while $\sim 10^6$
trajectories are needed to reduce relative deviation below 1\%. Note
that a small number of sampled trajectories also causes a systematic
error: the smaller the number of trajectories, the lower the averaged
$v_2$.

When using the equidistant sampling method instead of Monte Carlo, we divide the transverse plane into an equidistant grid, whose points are used as jet origins. Energy loss for each trajectory is
then weighted with the jet production probability at each
point, and summed up. As production probability, we used the binary
collision density evaluated using the optical Glauber model. In
Fig.~\ref{fig:SCANConvergence}, we see that, for already $\sim
10.000$ evaluated trajectories, the integral has converged within 1\%
of the estimated 'proper' value. This modification resulted in a more
than two orders of magnitude reduction of the execution time. We also
tested two hybrid variants: {\it i)} where trajectory origins were
randomly selected but directions equidistantly,
and {\it ii)} where production points were equidistantly selected, but directions randomly sampled. The convergence of the
two variants interpolated between the MC sampling and the equidistant sampling (Figs.~\ref{fig:MCConvergence}
and~\ref{fig:SCANConvergence}, respectively).

\subsection{Convergence test of different DREENA methods}

\begin{figure*}[h]
\centering
\epsfig{file=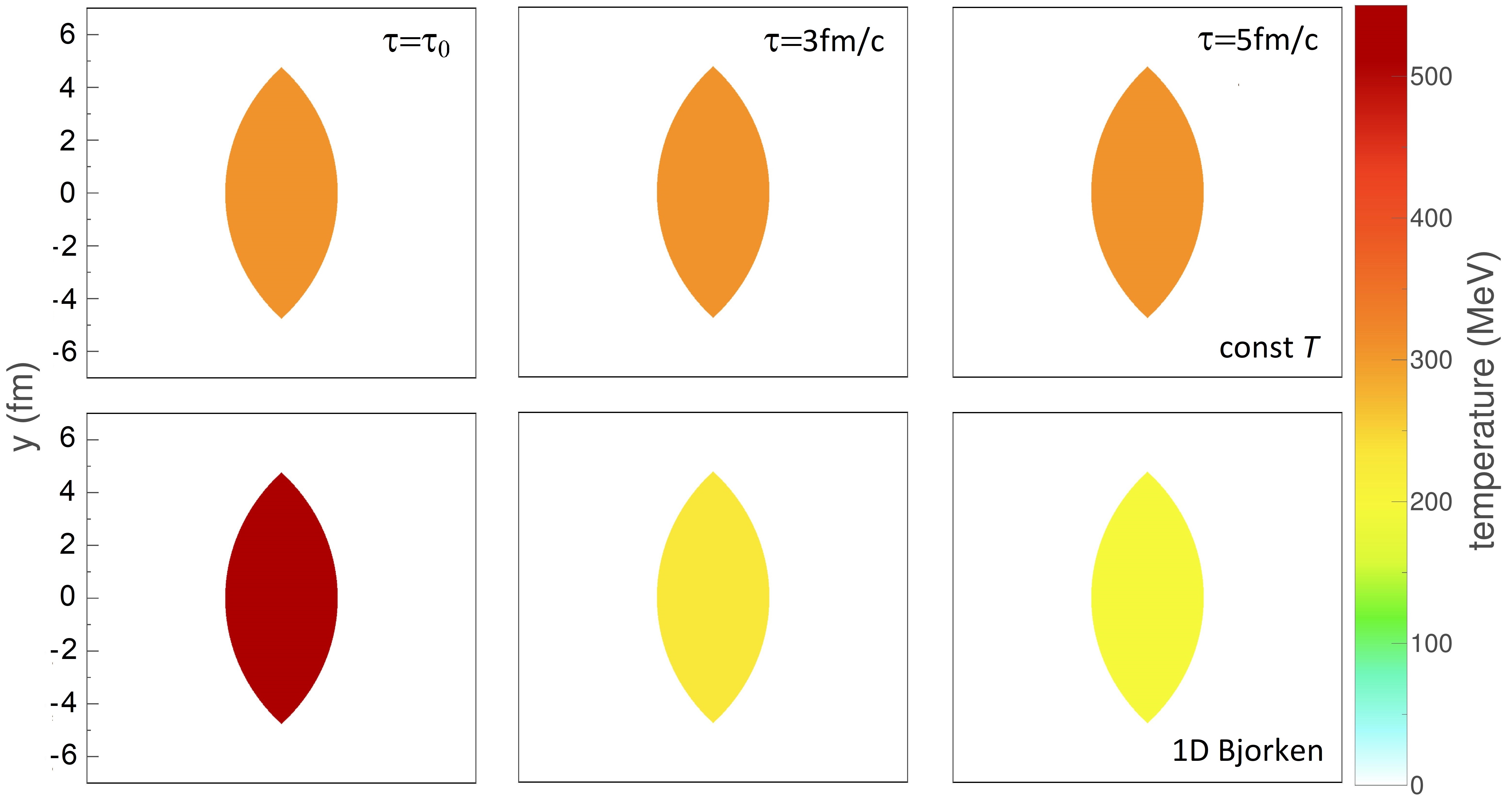,scale=0.2}
\vspace*{-0.2cm}
\caption{Temperature distribution (Pb + Pb collision, 30-40\%
  centrality, mid-rapidity) for constant temperature~\cite{DREENA-C} (first fow) and 1D Bjorken evolution~\cite{DREENA-B} (second row), at
  time (from left to right) $\tau = \tau_0,\, 3,$ and $5$
  fm/$c$, represented by colour mapping. For constant
  temperature approximation, $\tau_0 = 0$~fm. For 1D Bjorken
  approximation, $\tau_0 = 0.6$~fm.}
\label{fig:TempEvols_CB}
\end{figure*}

\begin{figure*}[h]
\centering
\epsfig{file=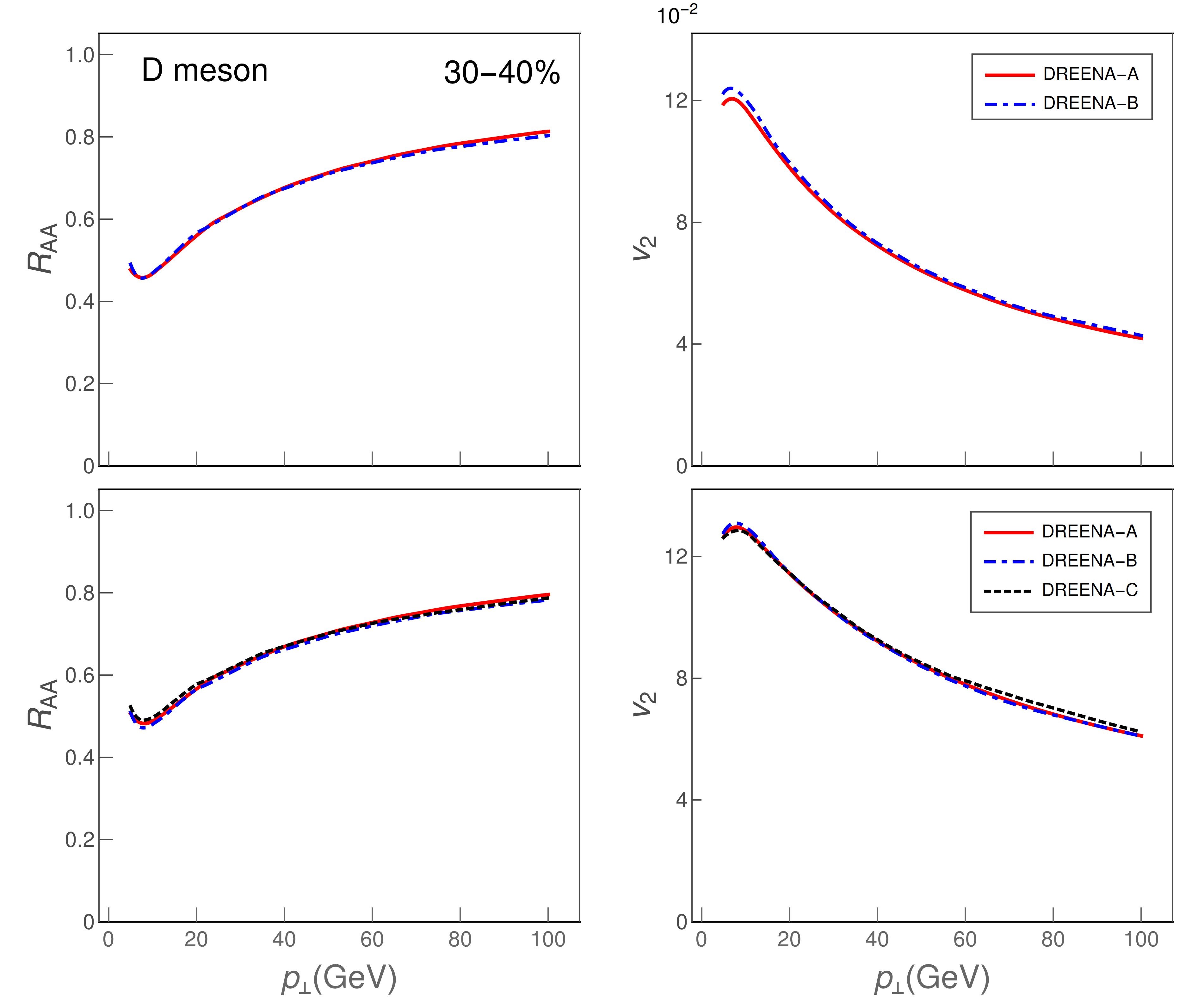,scale=0.375}
\vspace*{-0.2cm}
\caption{Comparison of different DREENA frameworks, for Bjorken medium
  evolution (upper panels) and for constant medium temperature
  approximation (lower panels), demonstrating inter-framework
  consistency. Upper panels show D meson $R_{AA}$ (left) and $v_2$
  (right) at 30-40\% centrality computed using DREENA-A (supplied with
  temperature profiles representing Bjorken evolution) and
  DREENA-B. Lower panels show the same observables, computed using all
  three DREENA frameworks, when applied to the same constant
  temperature medium. $M=1.2$~GeV, $\mu_M/\mu_E=0.5$.}
\label{fig:DREENALimits}
\end{figure*}

Finally, as a consistency check for DREENA-A, we compared its predictions with
DREENA-C and DREENA-B results. For this purpose, we generated artificial
$T$ profiles suitable for this comparison, illustrated in
Figure~\ref{fig:TempEvols_CB}. The results of the
DREENA-A and DREENA-B comparison, for $R_{AA}$ and $v_2$, are shown in
the upper panels of Figure~\ref{fig:DREENALimits}, respectively. Lower
panels of Figure~\ref{fig:DREENALimits} show the comparison of all
three frameworks on the hard-cylinder collision profile constant in
time (for this comparison, we modified the DREENA-B code to
remove temperature dependence on time). We see that all frameworks
lead to consistent results (up to computational precision),
supporting the reliability of the DREENA-A.

\section{Results}

To demostrate the utility of the DREENA-A approach, we generated
temperature profiles for Pb+Pb collisions at the full LHC energy
($\sqrt{\sNN} = 5.02$ TeV) and Au+Au collisions at the RHIC energy
($\sqrt{\sNN} = 200$ GeV) using three different initialisations of the
fluid-dynamical expansion.

\begin{figure*}[h]
\centering
\epsfig{file=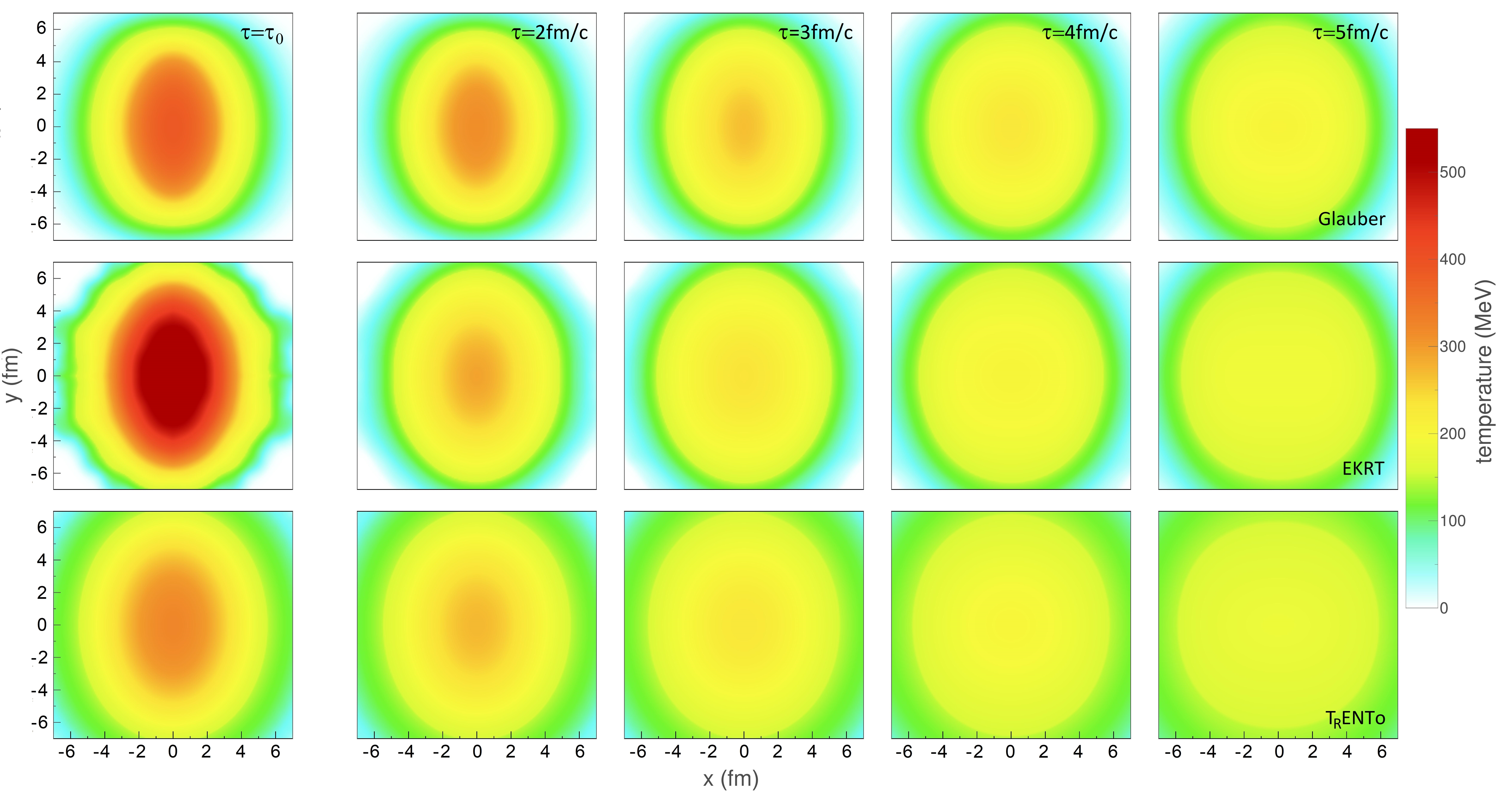,scale=0.2}
\vspace*{-0.2cm}
\caption{Temperature distribution (Pb + Pb collision, 30-40\%
  centrality, mid-rapidity 
  ) for different medium evolution models, at
  time (from left to right) $\tau = \tau_0,\, 2,\, 3,\, 4$ and $5$
  fm/$c$, represented by colour mapping. First row: 'Glauber', $\tau_0 = 1$~fm;
  second row: 'EKRT', $\tau_0 = 0.2$~fm; third row: 'T$_{\mathrm{R}}$ENTo',
  $\tau_0 = 1.16$~fm. Note that distributions in the first column correspond
  to different times.}
\label{fig:TempEvols}
\end{figure*}

First, we used optical Glauber initialisation at initial time $\tau_0
= 1.0$ fm without initial transverse flow. The evolution of the fluid
was calculated using a 3+1D viscous fluid code from
Ref.~\cite{Molnar:2014zha}. The parameters to describe collisions at
the LHC energy were tuned to reproduce the low-$\pT$ data obtained in
Pb+Pb collisions at $\sqrt{\sNN} = 5.02$ TeV~\cite{Stojku:2020wkh}. In
particular, shear viscosity over entropy density ratio was constant
$\eta/s = 0.12$, there was no bulk viscosity, and the equation of
state (EoS) parametrisation was $s95p$-PCE-v1~\cite{Huovinen:2009yb}.
For RHIC energy we used 'LH-LQ' parameters from
Ref.~\cite{Molnar:2014zha}, except that we used constant
$\eta/s = 0.16$.

Second, we used the EKRT initialisation~\cite{Eskola:1999fc,
  Paatelainen:2012at,Paatelainen:2013eea}, and evolved it using the
same code we used to evolve the Glauber initialisation, but
restricted to a boost-invariant expansion. In this case, the initial
time was $\tau_0 = 0.2$ fm, and parameters were the favoured values of
a Bayesian analysis of the data from Pb+Pb collisions at $\sqrt{\sNN}
= 2.76$ and $5.02$ GeV, and from Au+Au collisions at $\sqrt{\sNN} =
200$ GeV using the EoS parametrisation $s83s_{18}$~\cite{Niemi}. In
particular, there was no bulk viscosity and the minimum value of
temperature-dependent $\eta/s$ was 0.18.

Our third option was the T$_\mathrm{R}$ENTo
initialisation~\cite{Moreland:2014oya} evolved using the VISH2+1
code~\cite{Song:2007ux} as described in~\cite{Bernhard:2018hnz,
  Bernhard:2019bmu}. To describe collisions at LHC, parameters
were based on a Bayesian analysis of the data at the above mentioned
two LHC collision energies~\cite{Bernhard:2019bmu}, although the analysis
was done event-by-event, whereas we carried out the calculations using
simple event-averaged initial states. In particular, the calculation
included free streaming stage until $\tau_0 = 1.16$ fm, EoS based on
the lattice results by the HotQCD collaboration~\cite{HotQCD:2014kol},
and temperature-dependent shear and bulk viscosity coefficients with
the minimum value of $(\eta/s)_{\mathrm{min}} = 0.081$ and maximum of
$(\zeta/s)_{\mathrm{max}} = 0.052$.
For RHIC, we used the 'PTB' maximum a posteriori parameter values from
Ref.~\cite{JETSCAPE:2020mzn}, but
changed the temperature-dependent shear viscosity coefficient
$(\eta/s)(T)$ to a constant $\eta/s = 0.16$.

All these calculations lead to an acceptable fit to measured charged
hadron multiplicities, low-$\pT$ spectra, and
$\pT$-differential $v_2$ in $10-20$\%, $20-30$\%, $30-40$\%, and $40-50$\%
centrality classes. As we may expect, different initialisations and
initial times lead to a visibly different temperature evolution.
This is demonstrated in Fig.~\ref{fig:TempEvols} where we show the
calculated temperature distributions in collisions at the
LHC energy at various times. Even if
the initial anisotropy of the Glauber initialisation is lowest, later
in time, its anisotropy is largest, since the very early start of EKRT
initialisation, or the early free streaming of T$_\mathrm{R}$ENTo,
dilute the spatial anisotropy very fast. Similarly, the early start of EKRT
leads to a large initial temperature.

\begin{figure*}
\centering
\epsfig{file=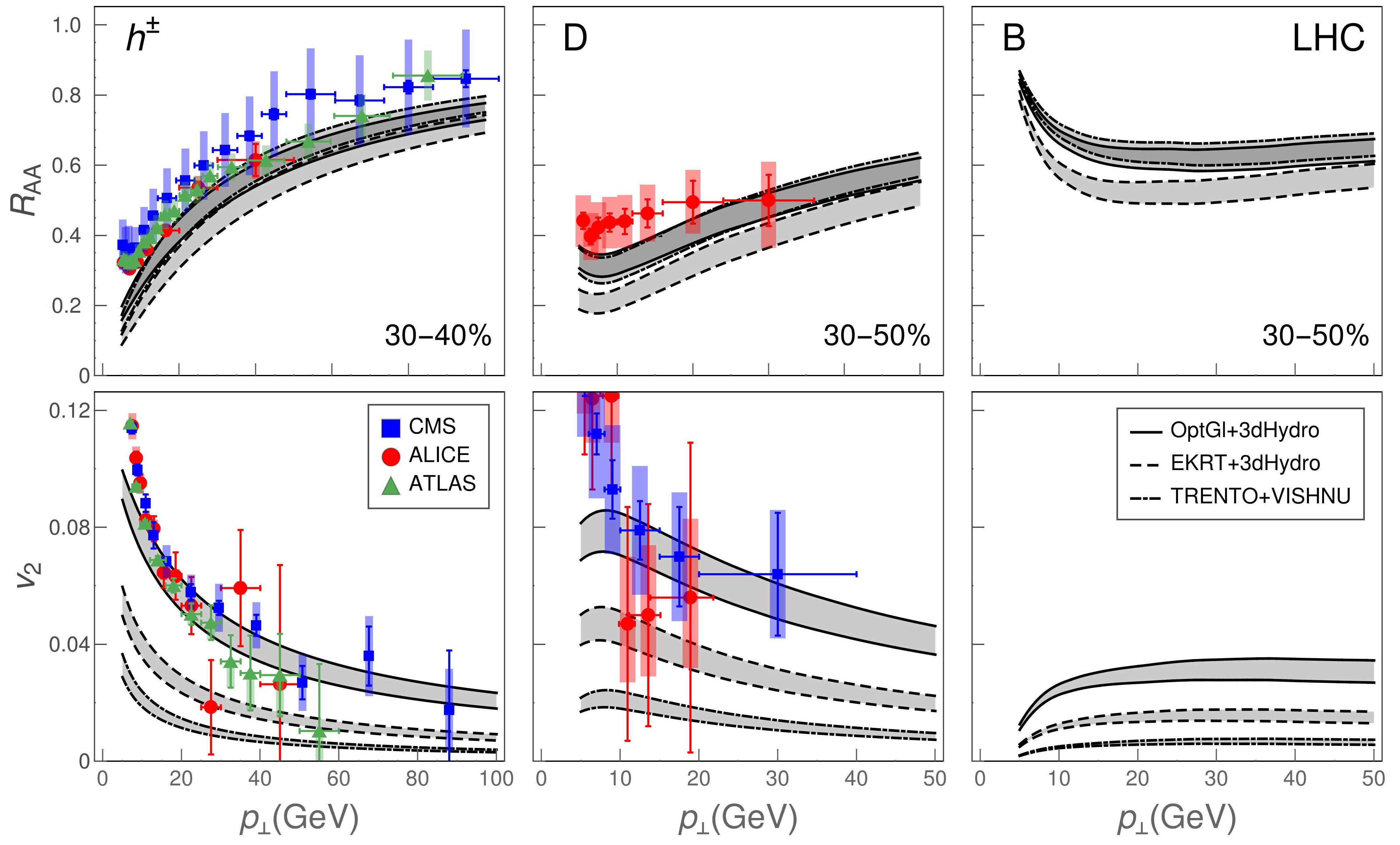,scale=0.42}
\vspace*{-0.2cm}
\caption{ DREENA-A $R_{AA}$ (top panels) and $v_2$ (bottom panels)
  predictions in Pb+Pb collisions at $\sqrt{\sNN} = 5.02$ TeV are
  generated for different models of QGP medium evolution (indicated in
  the legend). Charged hadron (left) predictions are generated for 30-40\% centrality, while D (middle) and B (right) meson predictions are generated for 30-50\% centrality region. For charged hadrons, the predictions are compared with the experimental data from CMS~\cite{CMS_CH_RAA,CMS_CH_v2},
  ALICE~\cite{ALICE_CH_RAA,ALICE_CH_v2} and
  ATLAS~\cite{ATLAS_CH_RAA,ATLAS_CH_v2}. For D mesons, the predictions are compared with ALICE~\cite{ALICE_D_RAA,ALICE_D_v2} and CMS~\cite{CMS_D_v2} data.
  The boundary of each gray
  band corresponds to $0.4<\mu_M/\mu_E<0.6$~\cite{Maezawa,Nakamura}.}
\label{Raav2}
\end{figure*}

\begin{figure*}
\centering
\epsfig{file=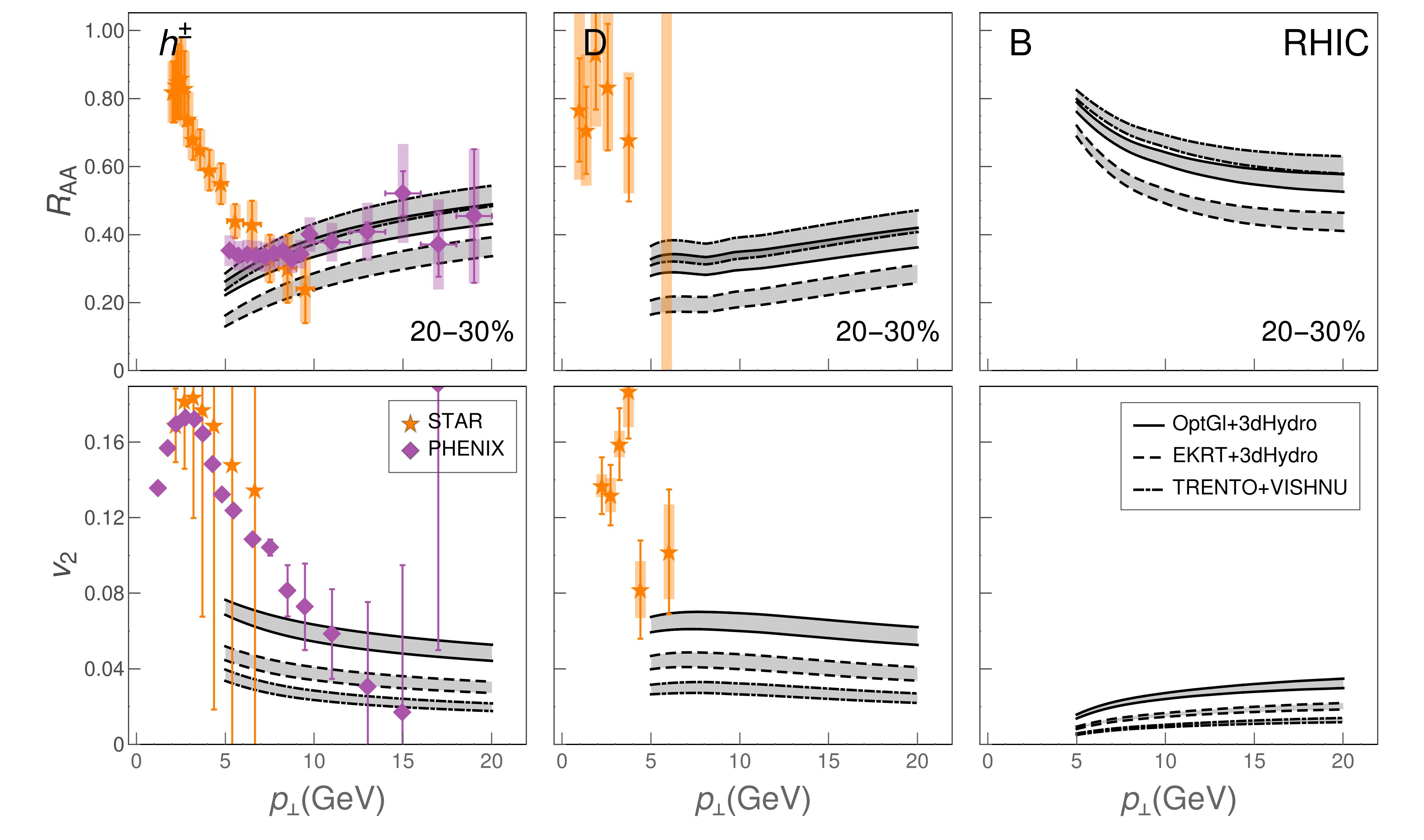,scale=0.42}
\vspace*{-0.2cm}
\caption{ DREENA-A $R_{AA}$ (top panels) and $v_2$ (bottom panels)
  predictions in Au+Au collisions at $\sqrt{\sNN} = 200$ GeV are
  generated for different models of QGP medium evolution (indicated in
  the legend). Charged hadron (left), D meson (middle) and B meson (right) predictions are generated for 20-30\% centrality region. The $h^\pm$ predictions are compared with $\pi^0$ data from PHENIX~\cite{PHENIX_PI_RAA,PHENIX_PI_v2} and $h^\pm$ data from STAR~\cite{STAR_CH_RAA,STAR_CH_v2} - note that for $v_2$ 10-40\% centrality data is shown for STAR. For D mesons, the predictions are compared with STAR~\cite{STAR_D_RAA,STAR_D_v2} data at 10-40\% centrality.
  The boundary of each gray
  band corresponds to $0.4<\mu_M/\mu_E<0.6$~\cite{Maezawa,Nakamura}.}
\label{Raav2_RHIC}
\end{figure*}

To test if these visual differences can be quantified through high-$\pT$
data at the LHC and RHIC, we used these profiles as an input to the DREENA-A to
generate high-$\pT$ $R_{AA}$ and $v_2$ predictions for charged
hadrons, D and B mesons. As can be seen in Figs.~\ref{Raav2} and~\ref{Raav2_RHIC}, both $R_{AA}$ and $v_2$
show notable differences for both experiments and all types of flavor. For example, 'EKRT' leads to the smallest
$R_{AA}$, as can be expected based on the largest temperature.
Similarly, the calculated high-$\pT$ $v_2$ depicts the same ordering
as the system anisotropy during the evolution: 'Glauber' leads to
the largest, and T$_\mathrm{R}$ENTo to the lowest $v_2$.
Consequently, the DREENA-A framework can differentiate between
temperature profiles by corresponding differences in
high-$\pT$ observables. Since the differences in evolution are due to
different initialisations, and different properties of the fluid (EoS
and/or dissipative coefficients), $R_{AA}$ and $v_2$ observables can
be used to provide further constraints to the fluid properties. We
note here that even low-$\pT$ data could be used to differentiate our
three evolution scenarios, but such analysis would require evaluating
$\chi^2$ or a similar measure of the quality of the fit, or computing
Bayes factors~\cite{JETSCAPE:2020mzn}. The high-$\pT$ observables, on the other hand, show clear differences visible by the naked eye.

Moreover, from Figs.~\ref{Raav2} and~\ref{Raav2_RHIC}, we see that
all types of flavor, at both RHIC and LHC, show apparent
sensitivity to differences in medium evolution, making them equally suitable
for exploring the bulk QGP properties with high-$\pT$ data. With the expected availability of precision data from the upcoming
high-luminosity experiments at RHIC and LHC (see e.g.,~\cite{sPHENIX,STAR_BUR,LHC_Run3}), the DREENA-A framework
 provides a unique opportunity for exploring the
bulk QGP properties. We propose that the adequate medium evolution should be able to
reproduce high-$\pT$ observables in both RHIC and LHC experiments
for different collision energies and collision systems, with reasonable accuracy. As demonstrated in this study, an equal emphasis should be given
to light and heavy flavor, as they provide a valuable independent constraint
for bulk medium evolution. Overall, DREENA-A provides a versatile tool to put large amounts of data generated at RHIC and LHC experiments to optimal use.

\section{Summary}

We here presented an optimised DREENA-A computational framework. The tool is based on state-of-the-art energy loss calculation and can
include arbitrary temperature profiles. This feature allows fully
exploiting different temperature profiles as the only input in the
framework. We showed that the calculated high-$\pT$ $R_{AA}$ and $v_2$
exhibit notable sensitivity to the details of the temperature profiles,
consistent with intuitive expectations
based on the profile visualisation. The DREENA-A framework
applies to different types of flavor, collision systems, and collision
energies. It can, consequently, provide an efficient and versatile QGP tomography tool for further constraining the bulk properties of this extreme form
of matter.

{\em Acknowledgements:}
We thank Marko Djordjevic, Bojana Ilic, and Stefan Stojku for useful
discussions. This work is supported by the European Research Council, grant
ERC-2016-COG: 725741, and by the Ministry of Science and Technological
Development of the Republic of Serbia. PH was also supported by the
program Excellence Initiative Research University of the University of
Wroc\l{}aw of the Ministry of Education and Science.

\end{document}